\documentclass[journal]{IEEEtran}

\usepackage{booktabs}
\usepackage{graphicx}
\usepackage{hyperref}

\usepackage[utf8]{inputenc}
\usepackage[inline]{enumitem}
\usepackage{xspace}  
\usepackage{listings}
\usepackage{balance}
\usepackage{tikz}
\usepackage{xcolor}
\usepackage{multirow}
\usepackage[table]{xcolor}
\usepackage[most]{tcolorbox}
\usepackage{float}
\usepackage{subcaption}
\newfloat{listing}{tbp}{lol}
\floatname{listing}{Listing}
\usepackage{booktabs,xcolor,colortbl}
\usetikzlibrary{positioning,shapes.geometric}
\newtheorem{definition}{Definition}
\usetikzlibrary{positioning,shapes.geometric}

\lstdefinelanguage{diff}{
  basicstyle=\ttfamily\scriptsize,
  breaklines=true,
  frame=single,
  numbers=none,
  captionpos=b,
  aboveskip=0.5\baselineskip,
  belowskip=0.5\baselineskip,
  showstringspaces=false,
  morecomment=[f][\color{red!70!black}]{-},
  morecomment=[f][\color{green!50!black}]{+},
  morecomment=[f][\color{blue!50!black}\itshape]{@@},
}

\newcommand{\data}{157\xspace}

\newcommand{\etal}{\emph{et~al\mbox{.}}\xspace}
\newcommand{\bigbag}{\textsc{BigBag}\xspace}

\begin{document}

\title{Agentic Generation of AST Transformation Rules for Fixing Breaking Updates}

\author{%
\begin{tabular}[t]{ccc}
  \textbf{Frank Reyes} &
  \textbf{Benoit Baudry}&
  \textbf{Martin Monperrus}\\[3pt]
  \small\textit{KTH Royal Institute of Technology} &
  \small\textit{Universit\'{e} de Montr\'{e}al} &
  \small\textit{KTH Royal Institute of Technology} \\
  \small Stockholm, Sweden &
  \small Montr\'{e}al, Canada &
  \small Stockholm, Sweden \\
  \small frankrg@kth.se &
  \small benoit.baudry@umontreal.ca &
  \small monperrus@kth.se
\end{tabular}%
}

\maketitle

\begin{abstract}
Modern software projects depend on third-party libraries that evolve continuously, introducing breaking API changes that prevent client code from compiling after a dependency update.
When the same library update breaks multiple projects, existing repair approaches generate project-specific patches that cannot be reused, requiring each affected project to be repaired independently.
We present \bigbag, an agentic framework that generates fixing transformations: structured, executable programs that encode the repair logic at the API level and transfer to any client broken by the same update.
We evaluate \bigbag on 157 compilation failure breaking dependency updates from the BUMP benchmark, across eight configurations combining four large language models and two AST transformation engines (Spoon and JavaParser).
The best configuration achieves a compilable transformation rate of 94.3\% and a fix rate of 78.6\%.
Generated transformations transfer across projects, achieving a cross-project fix rate of 33.3\% overall and 80\% or above for breaking updates where all clients invoke the affected API element uniformly.
These results show that agentic generation of reusable fixing transformations is a viable approach to scalable repair of breaking updates.
\end{abstract}


\IEEEpeerreviewmaketitle

\section{Introduction}

\IEEEPARstart{M}{odern} software systems heavily depend on third-party libraries managed through package managers such as Maven, npm, and PyPI~\cite{decan2019empirical}.
These libraries continuously evolve~\cite{8330214}, and not all updates preserve backward compatibility.
Keeping dependencies up to date is therefore an essential maintenance activity, and tools such as Dependabot and Renovate~\cite{mirhosseini2017can} automate the version-update step at scale.
Removing a method, altering a signature, or relocating a type between packages causes dependent client projects to fail to build after the update~\cite{dig2006apis,ochoa2022breaking}.
These events are called \emph{breaking dependency updates}, and they are common: Ochoa~\etal~\cite{ochoa2022breaking} show that incompatible changes affect Maven clients even in minor and patch-level releases that should be safe according to SemVer.

When multiple projects depend on the same library, the same breaking dependency update affects all of them, each requiring the same conceptual fix applied to different code.
Yet, existing breaking update repair approaches address this one project at a time, generating project-specific code patches that cannot be reused across clients~\cite{reyes2025byam,FruntkeLukas2025}.
This limits the scalability of existing repair approaches: every affected project must be repaired independently, even when the underlying breaking dependency update is identical.

We address this problem with \bigbag, a novel technique based on reusable code transformations. \bigbag exploits a structural property of breaking dependency updates: all clients broken by the same update are broken because of the same root cause. The core idea of \bigbag is to generate \emph{fixing code transformations}: structured, standalone, executable programs that traverse a client project's AST, locate the constructs affected by a breaking change, and rewrite them to conform to the new API.
Unlike project-specific patches, a fixing transformation encodes the repair logic at the API level; once it is generated from one affected client, it applies to any other client broken by the same breaking dependency update.

\bigbag generates those fixing transformations with a coding agent through a generate-apply-verify loop~\cite{yao2023react} in which:
1) the agent builds the transformation against an AST transformation engine, 
2) applies it to the broken client until success based on compiler feedback;
3) deploys the transformation over other client projects affected by the same breaking update.

To evaluate \bigbag, we consider real world breaking dependency updates from BUMP~\cite{10589737}, a curated benchmark of reproducible breaking updates in real Maven projects. We study eight configurations combining four large language models (LLMs) and two AST transformation engines.

Our evaluation on \data{} breaking dependency updates across eight (model, engine) configurations shows that coding agents can reliably generate compilable fixing transformations.
The best configuration reaches a compilable transformation rate of 94.3\% and a fix rate of 78.6\%.
The choice of AST transformation engine is a first-order determinant of success: across four models, the two engines differ by up to 17.8\% in the rate of compilable transformations generated: JavaParser works generally better.
Next, we check whether generated transformations transfer across projects. They do, but not perfectly, with a cross-project fix rate of 33.3\% overall and 80\% or above for breaking updates where all clients invoke the affected API element uniformly.
Transfer success depends on usage uniformity across clients.

To the best of our knowledge, \bigbag is the first approach to generate AST transformations for fixing breaking dependency updates.
\bigbag also provides the first evidence of how AST transformation engine choice impacts repair success.
The closest contribution is~\cite{Ramos2026SPELL}, which also generates LLM-based migration artifacts that transfer across projects affected by the same API change; the key difference is that \bigbag targets Java breaking dependency updates using structured AST transformations rather than structural pattern-matching scripts for Python API migrations.

To summarize, our contributions are:
\begin{itemize}
  \item \bigbag, an agentic framework that generates AST transformations for fixing breaking dependency updates, available at~\url{https://github.com/chains-project/bigbag}.
  \item An empirical study across \data{} real-world breaking dependency updates and eight (model, AST engine) configurations, measuring transformation well-formedness, effectiveness, and cross-project transferability. 
  This systematic experimental campaign demonstrates the feasibility of generating fixing transformations for breaking dependency updates.
  \item A characterization of failure modes, showing that compilation success is correlated to the AST transformation engine API complexity, and that transfer success is determined by usage uniformity across clients.
\end{itemize}

The remainder of this paper is organized as follows.
\autoref{sec:background} introduces breaking dependency updates, AST transformation engines, and coding agents.
\autoref{sec:approach} describes the \bigbag framework.
\autoref{sec:study} presents the experimental methodology.
\autoref{sec:results} reports the results for each RQ.
\autoref{sec:threats} discusses threats to validity.
\autoref{sec:rw} discusses related work, and \autoref{sec:conclusion} draws conclusions.

\section{Background}
\label{sec:background}

\subsection{Breaking Dependency Updates}
\label{bkg:breaking}

Third-party libraries continuously evolve, and some evolutions are not backward-compatible, e.g., by removing methods, altering signatures, or introducing checked exceptions. When client projects update a dependency to a new version, they fail to compile due to the resulting API incompatibilities.
These events are called \textit{breaking dependency updates}~\cite{10589737}.
We adopt the following definitions from previous work~\cite{reyes2025byam}:

\begin{definition}
\label{def:depupdate}
A \textbf{dependency update} is a change made in a build specification file where the version of a specific dependency is updated to a new version. In this paper, we focus on the build file in Java/Maven: \texttt{pom.xml}.
\end{definition}

\begin{definition}
\label{def:breaking}
A \textbf{breaking dependency update} is a dependency update $(L,\, v_i \!\to\! v_j)$ that introduces incompatibilities causing the Maven build to fail, where $L$ denotes the updated library, $v_i$ the previous version, and $v_j$ the new version.
\end{definition}

Existing tools such as Dependabot and Renovate automate the version update step but do not repair the resulting compilation failures~\cite{mirhosseini2017can,kula2018developers}.
Recent LLM-based approaches address this repair step by generating code patches for individual affected projects~\cite{reyes2025byam,FruntkeLukas2025}.
\bigbag instead generates structured, executable fixing transformations: a single transformation derived from one client applies to any other client affected by the same breaking change.

\subsection{AST Transformations}
\label{bkg:AST}

Abstract Syntax Tree (AST) transformation is a structured approach to source-code modification in which a program is parsed into a typed tree, algorithmically rewritten, and regenerated from the modified tree~\cite{1265817,10.1145/1111542.1111544}.
A key component of this process is the AST transformation engine, which controls transformation generation.

\begin{definition}
\label{def:astengine}
An \textbf{AST transformation engine} is a tool that executes an AST transformation: it parses source code into a typed tree, exposes an API for traversing and modifying tree nodes, resolves type and symbol information, and pretty-prints the rewritten tree back to source.
\end{definition}

Engines differ along several dimensions: the traversal abstraction they expose (e.g., visitor pattern vs.\ processor)~\cite{gamma1993design}, the depth of their type-resolution model (fully resolved vs.\ syntactic only), the amount of scaffolding required to express a transformation, and the fidelity with which they preserve formatting during regeneration.
Two Java AST transformation engines are used in this study: Spoon and JavaParser.
\emph{Spoon}~\cite{pawlak:hal-01169705} provides a \texttt{Processor<T>} abstraction that encapsulates traversal and transformation logic in a single class and ships an integrated, fully resolved type model, enabling semantically rich transformations.
\emph{JavaParser}~\cite{smith2017javaparser} offers a lighter \texttt{Visitor} API; it requires less scaffolding for syntactic changes but provides fewer semantic guarantees.
Together, these two engines span the expressiveness-versus-simplicity design space for Java source transformation tools.
Both provide the primitives needed to implement a fixing transformation for a breaking update.

We define a fixing transformation as follows:
\begin{definition}
\label{def:astrule}
A \textbf{fixing transformation} is an executable program that (i)~traverses the client project's AST, (ii)~locates the constructs affected by breaking API changes, (iii)~rewrites them to conform to the new API, and (iv)~regenerates the modified source files.
\end{definition}

\subsection{Coding Agents}
\label{bkg:agents}

LLM-based coding agents operate through an observe-plan-act-verify loop~\cite{yang2024swe}.
The agent reads source files and compiler output, formulates a repair strategy, writes or modifies code, executes a build, and iterates on compiler feedback until the build succeeds or an iteration budget is exhausted.
Coding agents provide file-system and shell-execution capabilities, enabling the agents to interact directly with real codebases.
This iterative, feedback-driven loop is necessary for transformation synthesis: generating a correct executable AST transformation requires repeated compilation and build verification that a single-pass LLM call cannot provide.

Existing agent-based approaches to breaking update repair, such as those of Reyes~\etal~\cite{reyes2025byam} and Fruntke and Krinke~\cite{FruntkeLukas2025}, generate instance-specific patches that target one client project and cannot be transferred to others.
\bigbag instead constrains the agent to produce a fixing transformation that encodes the repair logic independently of any particular client project, making it applicable to any client broken by the same dependency update.

\begin{figure*}[t]
  \centering
  \includegraphics[width=0.95\linewidth]{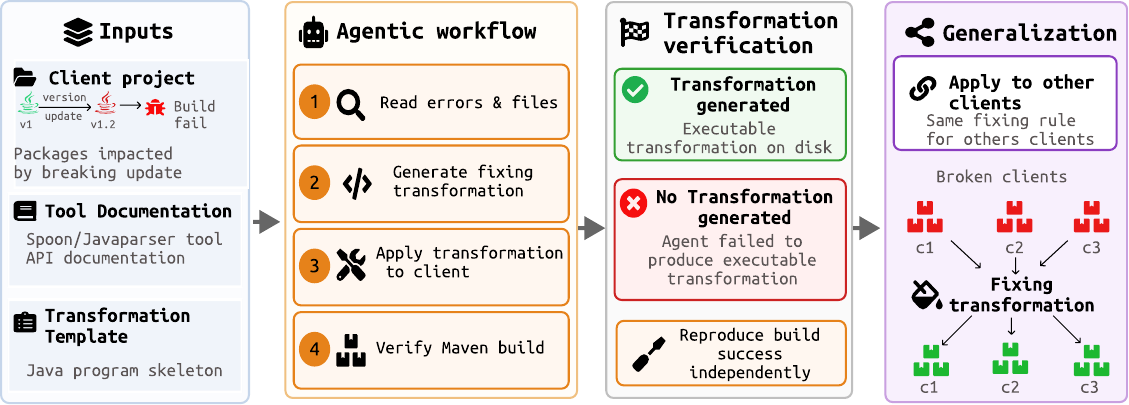}
  \caption{Overview of the \bigbag workflow.
  Step~1: \bigbag passes three inputs to the agent: the client project with a broken build, the AST engine API documentation, and a transformation template.
  Step~2: the agent executes a generate-apply-verify loop, generating a fixing transformation, applying it to the client sources, and triggering a Maven build; on failure, the agent revises and retries until the build succeeds or the iteration budget is exhausted.
  Step~3: \bigbag classifies each execution by whether a transformation was produced; for those that were, it verifies in isolation that the transformation alone resolves the compilation failure. All generated transformations are saved; only those passing verification are eligible for Step~4.
  Step~4: each eligible transformation is applied to all client projects affected by the same breaking change to assess cross-project transferability.}
  \label{fig:overview}
\end{figure*}

\section{\bigbag}
\label{sec:approach}

\bigbag is an agent-based workflow that generates reusable code transformations for fixing breaking dependency updates.
\autoref{fig:overview} provides an overview of the four-step process: input context assembly, transformation generation and application, transformation verification, and transformation generalization.

\begin{listing*}[t]
\begin{lstlisting}
public class Main {
  public static void main(String[] args) {
    Launcher launcher = new Launcher();
    launcher.addInputResource(args[0]);
    launcher.setSourceOutputDirectory(args[1]);
    CtModel model = launcher.buildModel();
    for (CtConstructorCall<?> c :
        model.getElements(new TypeFilter<>(CtConstructorCall.class))) {
      if (c.getType().getQualifiedName().equals(
              "org.yaml.snakeyaml.constructor.Constructor")
          && c.getArguments().isEmpty()) {
        CtExpression<?> arg = launcher.getFactory()
            .createCodeSnippetExpression("new org.yaml.snakeyaml.LoaderOptions()");
        c.addArgument(arg);
      }
    }
    launcher.prettyprint();
  }
}
\end{lstlisting}
\caption{Generated fixing transformation by GeminiCLI/Gemini-3.1-Pro for update of \texttt{snakeyaml} 1.33\,$\to$\,2.0 in \texttt{fluxtion}. The \texttt{Constructor} class no longer accepts a no-argument instantiation; the transformation traverses all constructor calls of that type and adds the required \texttt{LoaderOptions} argument.}
\label{list:example_rule_repair}
\end{listing*}

\subsection{Step 1: Input Context Assembly}
\label{sec:step1}

\bigbag assembles four inputs before invoking the agent: the client project under repair, the AST transformation engine documentation, the transformation template, and a composite fourth input comprising the new dependency API specification and its Javadoc.

\subsubsection{Client Project with Breaking Update}
The first input is the \emph{client project}, the Maven project whose build fails after updating a dependency from version~$v_i$ to version~$v_j$ (~\autoref{def:breaking}).
The project folder contains the full source tree and the manifest (\texttt{pom.xml}) that declares the updated dependency.
The agent identifies the affected source files autonomously by invoking \texttt{mvn compile} and reading the compiler logs, which report the exact files and lines where the API incompatibility occurs.

\subsubsection{AST Transformation Engine Documentation}
The second input is the \emph{AST transformation engine documentation} for the selected engine (Definition~\ref{def:astengine}), either Spoon or JavaParser (\autoref{bkg:AST}).
For each engine, we supply the standard Javadoc generated from the transformation library; the Javadoc archives for the versions used in our study contain between 1,500 and 1,900 HTML files per engine.
\bigbag{} provides the documentation directory as a file-system reference in the agent prompt.
The agent navigates and reads individual pages on demand rather than receiving the full documentation as a bulk context injection, to avoid potential context saturation; we discuss this design assumption in the Threats to Validity section.
This approach makes the exact API version of the transformation engine available to the agent at generation time, reducing reliance on pre-trained knowledge that may be outdated or incorrect for the target version.

\subsubsection{Transformation Template}
The third input is a \emph{transformation template}: a Maven project whose \texttt{pom.xml} declares the dependency on the assigned AST transformation engine and a Java file (\texttt{Main.java}) with an empty \texttt{main} method.
The template contains no pre-written transformation code.
It provides only the build configuration and package structure, so the agent can focus on generating the AST transformation logic without inferring dependency coordinates or build setup.
We provide one template per engine, each declaring the corresponding library as the sole Maven dependency; both templates are available in our replication package.

\subsubsection{Dependency API Specification and Javadoc}
The fourth input comprises two complementary documents that describe the public API surface of the updated dependency at version~$v_j$.
The \emph{API specification} is a Markdown file (\texttt{api-spec.md}) that lists all exported types of the dependency at version~$v_j$ with their complete public method signatures, including fully-qualified parameter and return types.
It serves as a compact reference for the agent to look up correct replacement signatures when the compiler reports an incompatible usage.
The \emph{dependency Javadoc} is the standard HTML documentation generated from the dependency source at version~$v_j$, providing prose descriptions, parameter semantics, return types, and usage examples for every public element.
The agent uses these two documents to identify correct replacement types, method signatures, and constructor parameters without relying on pre-trained knowledge that may reflect a different version of the library.
\bigbag{} provides the full description of what the new API \emph{is}; the agent derives the required fixing transformation by comparing the compiler errors in the client project against that documented surface.

\bigbag{} assembles these four inputs into the prompt (see \href{}{template on our repository}), which initializes the agent.

\subsection{Step 2: Transformation Generation and Application}
\label{sec:step2}

The agent begins by invoking \texttt{mvn compile} on the client project to collect the compilation errors.
It then reads the source files identified by those errors to locate the constructs affected by the incompatible API change.
From this analysis, the agent generates a standalone Java program that performs the fixing transformation (Definition~\ref{def:astrule}): it constructs an AST model of the affected sources, traverses the model to identify incompatible usages, and rewrites those usages to comply with the new API.

The structure of the generated program depends on the assigned AST engine.
For \emph{Spoon}, the agent produces a class that instantiates a \texttt{Launcher}, constructs a \texttt{CtModel}, and directly manipulates typed \texttt{CtElement} nodes.
For \emph{JavaParser}, the agent produces a class implementing the \texttt{Visitor} interface, traversing and rewriting \texttt{Node} objects; this engine requires less scaffolding than Spoon but provides fewer type-resolution guarantees, and is suited for transformations that operate on syntactic structure alone.

Listing~\ref{list:example_rule_repair} illustrates a Spoon fixing transformation generated for the \texttt{fluxtion} client project, where the \texttt{snakeyaml} 1.33\,$\to$\,2.0 update changed the \texttt{Constructor} class to require a \texttt{LoaderOptions} argument: the transformation locates every no-argument \texttt{Constructor()} call and adds \texttt{new LoaderOptions()} as the required argument.

The agent autonomously compiles each candidate fixing transformation and, if compilation succeeds, executes it against the client project's source tree.
During execution, the fixing transformation processes the source files of the client project, performs the AST traversal and rewriting, and writes the transformed output to disk.
If the fixing transformation fails to compile or raises an unhandled exception during execution, the agent treats this as a generation failure and re-enters the generation loop, using the compiler or runtime error as feedback for the next generation attempt.

If the fixing transformation executes without error, the agent autonomously triggers a Maven build of the client project.
The agent terminates autonomously when it determines the task is complete, without waiting for external confirmation that the overall repair succeeded.
This stopping condition means the agent may deliver a transformation that is syntactically malformed or that references undefined API elements.

\subsection{Step 3: Fixing Transformation Verification}
\label{sec:step3}

\bigbag classifies each execution at the conclusion of Step~2 into one of two mutually exclusive output classes based on a single criterion: whether the agent produced at least one executable fixing transformation (Definition~\ref{def:astrule}).

\textbf{Transformation Generated.}
An execution is classified as ``transformation-generated'' when the agent produces at least one executable fixing transformation, irrespective of whether the Maven build of the client project succeeds after transformation application.
If a transformation is generated, \bigbag re-applies the fixing transformation to its original client project and invokes the Maven build to verify that the fixing transformation resolves the compilation failure, outside of the agentic loop.
This verification is necessary because during Step~2 the agent may have applied additional modifications to the client project beyond those produced by the transformation itself, due to full autonomy.
Re-applying the transformation in isolation confirms that the transformation alone is sufficient to resolve the compilation failure.
All generated fixing transformations are saved; only those for which this verification build succeeds are eligible for Step~4.

\textbf{Fixing Transformation not generated.}
An execution is classified as transformation-not-generated when the agent does not produce an executable fixing transformation within its iteration budget.
This outcome may arise for several reasons:
\begin{itemize}
  \item the agent exhausts its iteration budget without producing compilable transformation code;
  \item the agent has modified the affected source files through direct edits without using a structured transformation.
\end{itemize}
In any of these cases, the execution produces no transferable artefact.

\subsection{Step 4: Cross-Project Transformation Transfer}
\label{sec:step4}

In step 4, \bigbag evaluates whether each transformation that passed the Step~3 verification transfers to client projects other than the one from which it was derived.
A breaking change $(L,\, v_i \!\to\! v_j)$ affects every client project that depends on library~$L$ and has updated from version~$v_i$ to version~$v_j$ (Definition~\ref{def:breaking}).
Because the incompatible usage pattern originates from the library update itself, all affected clients exhibit similar instances of the same incompatibility.

\bigbag uses the breaking change $(L,\, v_i \!\to\! v_j)$ associated with each transformation to identify all other client projects affected by the same library update: projects assumed to contain the same incompatible usage pattern introduced by the version transition from~$v_i$ to~$v_j$.
These projects form the \emph{validation set} of the transformation.

\bigbag applies each fixing transformation to every project in its validation set, executing the fixing transformation against the target project's source tree and invoking the build to determine whether the transformation resolves the compilation failure in that project.
\bigbag records the build outcome (success or failure) for each (transformation, target project) pair.
A transformation that successfully repairs one or more projects in its validation set provides evidence of generalizability: the code transformation captures a property of the API change itself rather than an artefact of the specific client project from which it was generated.

\subsection{Implementation}

\bigbag orchestrates all agent invocations, dataset preparation, and result collection in Java.
The generated fixing transformations are also Java programs, based either on the Spoon transformation engine or on the JavaParser transformation engine.

\textbf{Model selection.}
We integrate four frontier code-generation models: GPT-5.4-mini~\cite{gpt54minis} (OpenAI), Qwen3-30B~\cite{qwen330b} (Alibaba), DeepSeek-v3.2~\cite{liu2025deepseek} (DeepSeek~AI), and Gemini-3.1-Pro~\cite{gemini31pro} (Google DeepMind).
All rank among the top models on the SWE-bench Verified leaderboard at the time of the study.\footnote{\url{https://www.swebench.com}}
This selection provides one model per vendor (controlling for pretraining corpus effects~\cite{FruntkeLukas2025,reyes2025byam}), mixes proprietary and self-hostable open-weight models (enabling replication without commercial subscriptions), and spans nearly two orders of magnitude in experimental cost (\$60 for Qwen3-30B to \$2{,}600 for Gemini-3.1-Pro).
All four support multi-turn tool use required by the generate-apply-verify loop (\autoref{sec:step2}).

\textbf{Agent scaffolds.}
We use two coding-agent scaffolds: OpenCode~v1.3.0\footnote{\url{https://github.com/opencode-ai/opencode}} for GPT-5.4-mini, Qwen3-30B, and DeepSeek-v3.2, and GeminiCLI~v0.22.5\footnote{\url{https://github.com/google-gemini/gemini-cli}} for Gemini-3.1-Pro.
Both scaffolds expose file-system read/write and shell-execution capabilities, enabling the agent to invoke \texttt{mvn compile} (Apache Maven~v3.9.2), read compiler output, and write transformed source files within the same feedback loop.
Each agent execution runs in an isolated environment; build reproducibility is verified by re-running the verification step in Step~3 (\autoref{sec:step3}) outside the agentic loop.

\textbf{AST transformation engines.}
We use two transformation engines: Spoon~v11.2.1~\cite{pawlak:hal-01169705} and JavaParser~v3.27.1~\cite{smith2017javaparser}.
The agent is provided with a template \texttt{pom.xml} that pins the library version used in our study. The agent is also given the capability to read the matching Javadoc archive for the exact version under evaluation.

\section{Experimental Methodology}
\label{sec:study}

\newcommand\rqbuildsuccess{\textbf{(Transformation well-formedness):}
To what extent can a coding agent generate well-formed, compilable fixing transformations for breaking dependency updates?}

\newcommand\rqrulefix{\textbf{(Transformation Effectiveness):}
To what extent do the agent-generated fixing transformations successfully fix breaking dependency updates?}

\newcommand\rqgeneralization{\textbf{(Transformation Generalizability):}
To what extent can a fixing transformation generated for a specific breaking change fix other client projects affected by that same breaking change?}

\subsection{Datasets}
\label{study_subjects}

We base our study on BUMP~\cite{10589737}, a benchmark of 571 reproducible breaking dependency updates in Maven projects.
We analyze only the subset of breaking dependency updates that are classified as \texttt{COMPILATION\_FAILURE}, excluding dependency-resolution failures and test failures.
This filtering retained \data breaking dependency updates representing API-level breakage across 69 distinct client projects and 70 distinct third-party libraries.
Of the 69 client projects, 28 (40.6\%) contribute more than one breaking update across consecutive versions.
Our filtered dataset and replication scripts are available in our replication package~\footnote{\url{https://github.com/chains-project/bigbag}}.

The \data{} breaking dependency updates are distributed across three semantic-versioning types: 3 PATCH version updates (1.9\%), 63 MINOR version updates (40.1\%), and 91 MAJOR version updates (58.0\%).
The 91 major-version updates dominate (58.0\%), consistent with the expectation that major releases carry API-breaking changes by design~\cite{ochoa2022breaking}.
The 63 minor-version updates confirm prior findings that incompatible changes occur even in releases that should be backwards compatible per SemVer~\cite{ochoa2022breaking}, including patch-level updates.

\begin{figure}[t]
  \centering
  \includegraphics[width=\columnwidth]{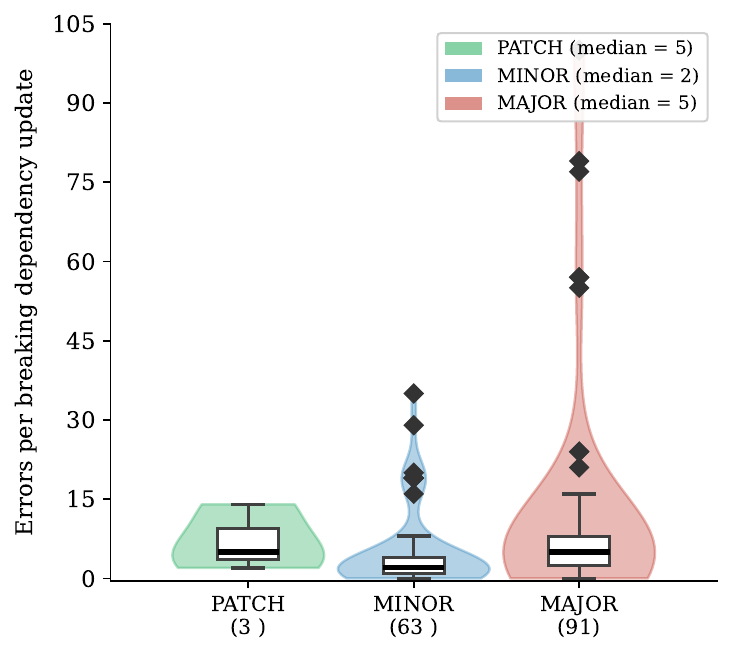}
  \caption{Distribution of compilation error counts per breaking updates by SemVer type across the \data{} study subjects.}
  \label{fig:dataset-violin}
\end{figure}

Figure~\ref{fig:dataset-violin} shows the distribution of compilation errors per breaking dependency update across all semver types.
MAJOR breaking updates have a median of ~5 compilation errors, with a maximum of ~100.
MINOR breaking updates have fewer problems (median~=~2, max~=~35).

\subsection{Research Questions}

In this section, we present the three research questions guiding our experiments.

\begin{enumerate}[label=\textbf{RQ\arabic*}, ref=RQ\arabic*]
 \item \label{rq:exploration}{\rqbuildsuccess}
 
 Executable transformation generation is a prerequisite for the approach to work at scale. An agent that produces no transformation has failed differently from one whose generated transformation does not compile.
 Distinguishing the two failure modes pinpoints where the agent's weaknesses lie.

 \item \label{rq:rulefix}{\rqrulefix}
 
 A syntactically valid transformation may still leave the project in a broken state if it targets the wrong construct or applies an incorrect rewrite. Measuring the fix rate quantifies this gap between transformation generation and repair effectiveness.

 \item \label{rq:generalization}{\rqgeneralization}

 A fixing transformation that applies only to the project it was generated for has limited practical value. Cross-project reuse is the property that makes the approach scalable in real dependency ecosystems.
\end{enumerate}

\subsection{Methodology for RQ1}
\label{meth:rq1}

RQ1 evaluates the extent to which \bigbag generates well-formed, compilable fixing transformations for breaking dependency updates.
We apply each of the four model configurations of \bigbag, described in ~\autoref{sec:approach}, to each of the \data breaking dependency updates described in ~\autoref{study_subjects}.
For each update, each model configuration is invoked independently once per AST transformation library, yielding $4 \times 2 \times \data$ total executions across eight (model, engine) combinations.
Results are stratified by semantic versioning type (PATCH, MINOR, MAJOR) to characterize agent behavior across the version categories in the dataset, as described in ~\autoref{study_subjects}.

For each execution, we first determine whether the agent produced a valid AST-based fixing transformation.
An output is considered invalid, and classified as a \emph{Transformation Bypass}, if one of the following conditions holds: (1) it uses string matching instead of AST operations; (2) it bypasses source-code transformation by modifying the \texttt{pom.xml} file; or (3) it does not use either of the two target AST libraries (Spoon or JavaParser).
Transformation Bypass outputs are divided into two sub-patterns: \emph{AST pretense}, in which the agent imports the target library but implements the transformation via \texttt{String.replaceAll()} or \texttt{java.util.regex.Pattern}; and \emph{strategy bypass}, in which the agent edits source files through \texttt{java.io.File} calls or modifies the \texttt{pom.xml} version string, instead of code transformation.
For each output that passes the first check, we then record whether it compiles without errors when built as a standalone Java program against the assigned library.
An output that passes the first check but fails to compile is classified as a \emph{compilation failure}.
An output that passes both checks is a \emph{compilable rule}, the successful outcome for RQ1.
Note that compilation of the fixing transformation is distinct from compilation of the client project, which is the subject of RQ2.

We use one key metric for RQ1: the \emph{Compilable Rule Rate} (CRR).
Let $E$ denote the set of all executions, $G \subseteq E$ the subset producing a valid AST-based fixing transformation, and $C \subseteq G$ the subset of those transformations that compile without errors.
CRR is the proportion $|C| / |E|$, reported per (model, engine, semver type) combination.

\subsection{Methodology for RQ2}
\label{meth:rq2}

RQ2 evaluates whether the compilable fixing transformations produced in RQ1 successfully repair the breaking update in the client project.
We take each transformation in $C$ (the subset that compiled without errors, as defined in \autoref{meth:rq1}), apply it to the corresponding client source files, and execute \texttt{mvn test} in the original Maven configuration.
We consider a transformation effective if \texttt{mvn test} exits without error, meaning the project compiles and all tests pass after the transformation is applied.

We address RQ2 with the \emph{Fix Rate} ($\mathrm{FR}$) metric: the proportion $|S|/|C|$, where $S \subseteq C$ is the set of compilable transformations for which \texttt{mvn test} succeeds, reported per (model, AST transformation engine, semver level) combination.
We use the same PATCH/MINOR/MAJOR stratification as RQ1.
Pairwise Fix Rate comparisons use Fisher's exact test; percentage-point differences serve as the effect size measure, consistent with the analysis in \autoref{meth:rq1}.

For transformations in $C$ that do not achieve Build Success, we classify the failure using the error categories defined in BUMP~\cite{10589737}.
We distinguish three observable failure modes: \emph{Client Compilation failure} (the client fails to compile after the transformation is applied); \emph{Rule execution error} (the transformation crashes at runtime or the build fails due to environmental factors such as Java version incompatibility or unresolvable transitive dependencies); and \emph{Test failure} (the client compiles but at least one test fails).
Client Compilation failure and Rule execution error indicate that the transformation likely targets the wrong construct, applies an incorrect rewrite, or misses a call site; Test failure indicates that the transformation resolves the compilation error but introduces behavioral changes.

\subsection{Methodology for RQ3}
\label{meth:rq3}

With this RQ, we evaluate the extent to which a fixing transformation produced by the agent for one client project, the \emph{seed project}, generalizes to additional client projects affected by the same breaking dependency update.
If a fixing transformation is well-formed, it operates at the API-usage level and should in principle apply to any client that uses the affected API element; this RQ tests that premise.
We restrict the transfer analysis to the best-performing configuration from RQ2, as it produces the largest pool of Build~Success transformations and therefore maximizes the number of transfer opportunities available for analysis.
Transferability is a property of the API change structure, not of the generating configuration. Any correct transformation for a given update encodes the same API-level rewrite, so the patterns observed generalise to other configurations that produce correct transformations.

\textbf{Input selection.}
We focus on the breaking dependency updates for which the best-performing configuration of RQ2 produces a fixing transformation.
For each such update, defined by a library $L$ upgrading from version $v_i$ to $v_j$, we use the GitHub Search API to identify open-source Maven projects that declare $L$ version $v_i$ as a dependency.
Each breaking dependency update identifies a specific API element (a method, class, or field) whose signature or availability changed between $v_i$ and $v_j$.
We search each candidate project's source code for the fully-qualified name of that element and discard projects in which no such usage appears.
Breaking updates for which all candidate projects are discarded by the above criteria are excluded from the transfer analysis.

\textbf{Failure verification.}
For each candidate project, we replace $v_i$ with $v_j$ in the \texttt{pom.xml} and trigger a Maven build.
We accept the project as a verified transfer target if two conditions hold: the build fails, and the reported compilation errors reference the same API elements that broke in the seed project, not merely any error from the same dependency upgrade.
This criterion indicates that the failure originates from the same incompatibility, not from an unrelated configuration issue.
This step yields 16 qualifying breaking dependency updates and 129 verified transfer targets.

\textbf{Transformation transfer.}
For each verified transfer target, we apply the effective fixing transformation from RQ2 without modification and trigger a Maven build.

\begin{figure}[t]
  \centering
  \includegraphics[width=0.60\columnwidth]{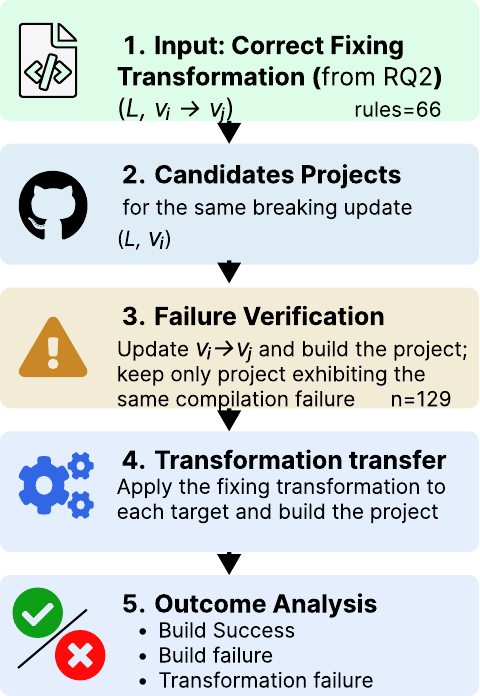}
  \caption{Methodology for RQ3: each effective transformation from RQ2 is applied
    without modification to other software packages affected by the same breaking
    dependency update; this enables us to compute the
    Cross-Project Fix Rate.}
  \label{fig:methodology-rq3}
\end{figure}

We answer RQ3 through the \emph{Cross-Project Fix Rate}: the proportion of verified transfer targets that achieve a successful Maven build after transformation application.
A non-fixed outcome occurs when the transformation executes without error but the client build remains failing, indicating that the transformation does not cover all API usages present in the external project.
\autoref{fig:methodology-rq3} illustrates this pipeline.

\section{Experimental Results}
\label{sec:results}

\subsection{RQ1: Well-Formedness of Generated AST Transformations}
\label{RQ1}

\begin{table*}[t]
  \centering
  \caption{Repair Well-formedness: number of compilable fixing transformations generated per semantic versioning level across all eight \bigbag\ configurations ($N{=}157$ per configuration). The \textbf{Total} row reports the Compilable Rule Rate ($\mathrm{CRR} = |C|/|E|$). Sp\,=\,Spoon; Jp\,=\,JavaParser. $^{*}$: highest count within that row.}
  \label{tab:rq1-compiled}
  \resizebox{\textwidth}{!}{%
  {\setlength{\tabcolsep}{4pt}
  \begin{tabular}{l c c c c c c c c}
    \toprule
    \textbf{Scaffold} & \multicolumn{6}{c}{\textbf{OpenCode}} & \multicolumn{2}{c}{\textbf{Gemini}} \\
    \cmidrule(lr){2-7}  \cmidrule(lr){8-9}
    \textbf{Model} & \multicolumn{2}{c}{\textbf{GPT-5.4-mini}} & \multicolumn{2}{c}{\textbf{Qwen3-30B}} & \multicolumn{2}{c}{\textbf{DeepSeek-v3.2}} & \multicolumn{2}{c}{\textbf{Gemini-3.1-Pro}} \\
    \cmidrule(lr){2-3}  \cmidrule(lr){4-5}  \cmidrule(lr){6-7}  \cmidrule(lr){8-9}
    \textbf{Engine} & \textbf{Sp} & \textbf{Jp} & \textbf{Sp} & \textbf{Jp} & \textbf{Sp} & \textbf{Jp} & \textbf{Sp} & \textbf{Jp} \\
    \midrule
    \rowcolor{gray!10} \textsc{patch} ($n$=3) & 2 & 2 & 2 & \textbf{3}$^{{*}}$ & \textbf{3}$^{{*}}$ & \textbf{3}$^{{*}}$ & 1 & 1 \\
    \textsc{minor} ($n$=63) & 46 & 56 & 32 & 36 & 51 & \textbf{60}$^{{*}}$ & 58 & 33 \\
    \rowcolor{gray!10} \textsc{major} ($n$=91) & 65 & 83 & 45 & 46 & 72 & \textbf{85}$^{{*}}$ & 44 & 50 \\
    \midrule
    \textbf{Total} ($n$=157) & 113/157 (72.0\%) & 141/157 (89.8\%) & 79/157 (50.3\%) & 85/157 (54.1\%) & 126/157 (80.3\%) & \textbf{148/157 (94.3\%)}$^{{*}}$ & 103/157 (65.6\%) & 84/157 (53.5\%) \\
    \bottomrule
  \end{tabular}%
  }%
  }%
\end{table*}

We evaluate the ability of each of the eight \bigbag\ configurations to produce a compilable fixing transformation for each of the \data{} breaking dependency updates.
Table~\ref{tab:rq1-compiled} presents the number of compilable transformations produced per semantic versioning level; the Total row also reports the Compilable Rule Rate ($\mathrm{CRR} = |C|/|E|$) for each configuration.
CRR ranges from 50.3\% (OpenCode/Qwen3-30B/Spoon) to 94.3\% (OpenCode/DeepSeek-v3.2/JavaParser).
Most pairwise inter-model CRR differences are statistically significant (Fisher's exact test, p~$<$~0.05); three pairs are statistically indistinguishable: DeepSeek-v3.2 and GPT-5.4-mini on both engines (p~=~0.21; p~=~0.11), GPT-5.4-mini and Gemini-3.1-Pro on Spoon (p~=~0.27), and Qwen3-30B and Gemini-3.1-Pro on JavaParser (p~=~1.00).
These pairs, though indistinguishable in compilability, diverge substantially in repair effectiveness as we shall see in RQ2.
All configurations produce compilable transformations for at least half of all breaking updates, and the best configuration (OpenCode/DeepSeek-v3.2/JavaParser) succeeds in 94.3\% of cases.

\textbf{Effect of the AST transformation engine.}
JavaParser yields a higher CRR than Spoon for all four models.
The gap is largest with GPT-5.4-mini ($+$17.8~pp) and DeepSeek-v3.2 ($+$14.0~pp).
This is because JavaParser's API is simpler; for example, the \texttt{Visitor} API requires a single method override per node type.
Spoon's \texttt{Processor<T>}, by contrast, requires instantiating a \texttt{Launcher}, building a \texttt{CtModel}, and implementing processor logic in a separate class, providing more surface area for \emph{Transformation Bypass} outputs (\autoref{meth:rq1}).
Qwen3-30B is the exception: its CRR gap between engines is negligible (54.1\% vs.\ 50.3\%, p~=~0.57, not significant), as its failures are dominated by compilation errors that arise independently of engine choice.
Gemini-3.1-Pro is the only model which generates more compilable Spoon transformations than JavaParser ones (65.6\% vs.\ 53.5\%, p~=~0.04).
Inspection of Gemini's agent traces reveals the agent bypasses the target AST library on JavaParser configurations more often than on Spoon, generating code that the Transformation Bypass check defined in \autoref{meth:rq1} rejects.
These results indicate that reducing API complexity in the transformation framework can be as effective as model choice in improving CRR. %

\textbf{Semver stratification.}
As expected, CRR is lower for MAJOR updates than for MINOR updates in six of eight configurations, consistent with MAJOR updates having higher essential complexity. The two exceptions are OpenCode/GPT-5.4-mini/JavaParser (MAJOR: 91.2\%, MINOR: 88.9\%) and GeminiCLI/Gemini-3.1-Pro/JavaParser (MAJOR: 54.9\%, MINOR: 52.4\%), both showing negligible differences.
Notably, compilation failures on MINOR updates confirm that version numbers alone are an unreliable signal of repair difficulty, as already discussed for \autoref{fig:dataset-violin} and previous work ~\cite{ochoa2022breaking}.
The three PATCH updates are too few for reliable comparison.

\textbf{Representative trajectory.}
Listing~\ref{list:example_rule_repair} illustrates a successful trajectory (Spoon, GeminiCLI/Gemini-3.1-Pro).
The agent invokes \texttt{mvn compile}, identifies \texttt{Constructor()} as the incompatible call site, and reads the \texttt{snakeyaml} 2.0 API documentation to locate the required \texttt{LoaderOptions} argument.
It then generates a fixing transformation that adds \texttt{new LoaderOptions()} to every no-argument \texttt{Constructor()} instantiation in the client source tree.
The transformation compiles on the first attempt; the agent applies it and confirms a successful Maven build.
This trajectory illustrates the self-correcting loop that underpins successful cases: the agent uses compiler feedback to locate the API change and API documentation to identify the replacement, completing the repair in a single iteration.

\begin{figure}[t]
  \centering
  \begin{subfigure}[b]{0.49\linewidth}
    \centering
    \includegraphics[width=\linewidth]{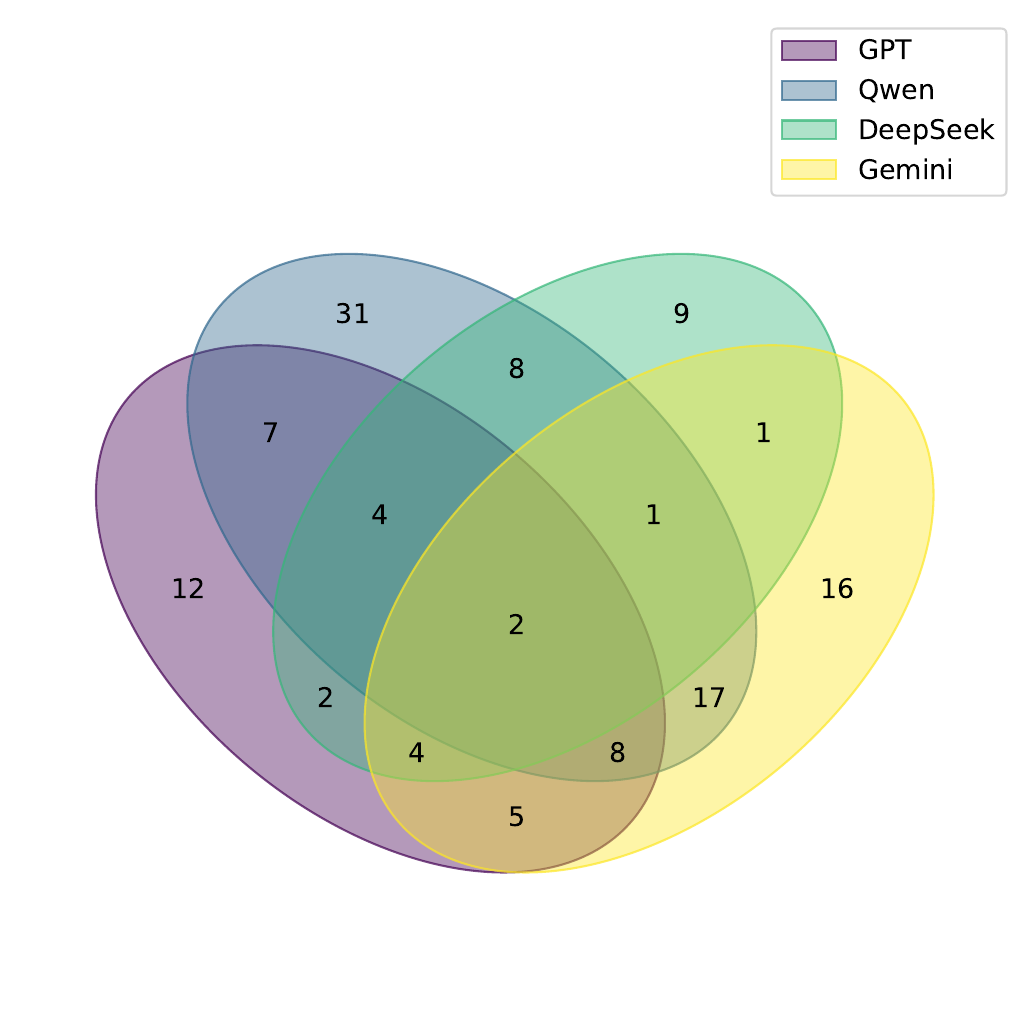}
    \caption{Spoon}
    \label{fig:venn-spoon}
  \end{subfigure}
  \hfill
  \begin{subfigure}[b]{0.49\linewidth}
    \centering
    \includegraphics[width=\linewidth]{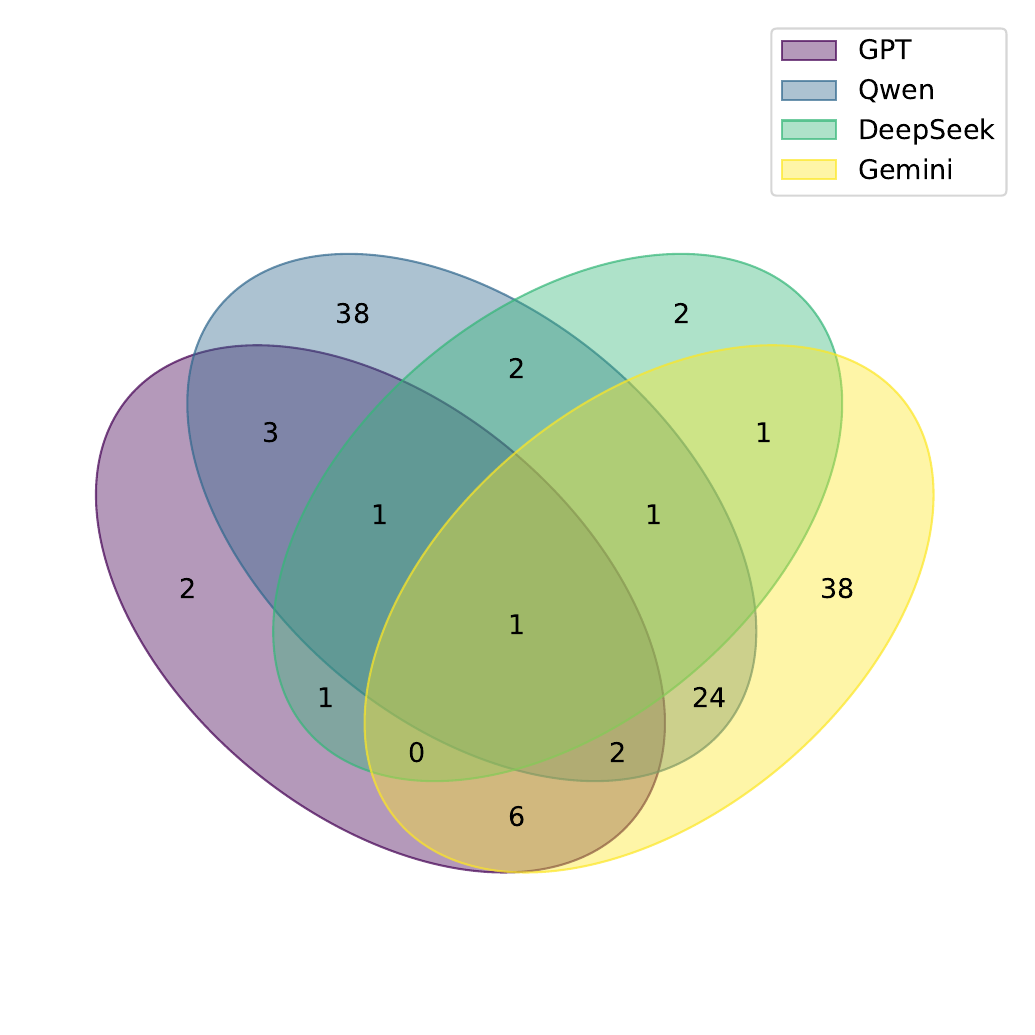}
    \caption{JavaParser}
    \label{fig:venn-javaparser}
  \end{subfigure}
  \caption{RQ1: non-compilable fixing transformations per model combination. Each region shows breaking updates where exactly that subset of models failed to produce a compilable fixing transformation ($N{=}157$).}
  \label{fig:venn-rq1}
\end{figure}

\textbf{Cross-model coverage.}
Figure~\ref{fig:venn-rq1} is a four-set Venn diagram partitioning all \data{} breaking updates by the subset of models that fails to produce a compilable fixing transformation on each engine.
With Spoon, 2 breaking updates fail with all four models simultaneously.
OpenCode/Qwen3-30B has the most exclusive failures (31), followed by GeminiCLI/Gemini-3.1-Pro (16), OpenCode/GPT-5.4-mini (12), and OpenCode/DeepSeek-v3.2 (9).
The dominant pairwise overlap is OpenCode/Qwen3-30B and GeminiCLI/Gemini-3.1-Pro (17 co-failures), with OpenCode/GPT-5.4-mini and OpenCode/Qwen3-30B adding a further 7 co-failures.
This indicates that Spoon's more complex API introduces failure sources shared across all models, not only the weakest configurations.
With JavaParser, failures concentrate more narrowly: only 1 breaking update fails with all four models simultaneously.
OpenCode/Qwen3-30B and GeminiCLI/Gemini-3.1-Pro each exclusively fail on 38 JavaParser breaking updates, yet through distinct mechanisms: OpenCode/Qwen3-30B's failures are compilation errors, while GeminiCLI/Gemini-3.1-Pro's are \emph{Transformation Bypass} outputs.
The largest pairwise overlap is the 24 JavaParser breaking updates in the OpenCode/Qwen3-30B and GeminiCLI/Gemini-3.1-Pro intersection, a co-failure between configurations that fail through opposite mechanisms.
These complementary failure patterns indicate that a multi-model ensemble accepting the first compilable rule across all four configurations would achieve 99.4\% CRR on JavaParser and 98.7\% on Spoon.
This reduces unresolved breaking updates to only those where all four models fail simultaneously: 1 on JavaParser and 2 on Spoon.

\textbf{Hardest cases.}
Three breaking updates never yield a compilable fixing transformation across all configurations: one with JavaParser and two with Spoon.
The \texttt{snakeyaml} 1.17$\,{\to}\,$2.0 update in \href{https://github.com/chains-project/bump/blob/main/data/benchmark/1e17e176460ab4283e463e62fece844d341da7f0.json}{\texttt{takari/polyglot-maven}} introduces 271 compilation errors; confronted with this scale, all models either hallucinate non-existent API classes, generate structurally malformed processors, or abandon the AST framework entirely.
The \path{flyway-core} 3.2.1$\,{\to}\,$9.15.0 update in \href{https://github.com/chains-project/bump/blob/main/data/benchmark/1d43bce1de6a81ac017c233d72f348d3c850299e.json}{\texttt{NemProject/nem}}, spanning 1{,}551 breaking changes because of a whopping six major version bump, clearly too extreme even for the best agents.
The \path{org.jenkins-ci:acceptance-test-harness} MINOR update in \href{https://github.com/chains-project/bump/blob/main/data/benchmark/7e8c62e2bb21097e563747184636cf8e8934ce98.json}{\texttt{jenkinsci/code-coverage-api-plugin}}, with only 7 breaking changes, fails using Spoon because replacing the affected expression requires \texttt{CtFactory} type inference patterns that none of the four models applies correctly.
These cases illustrate two distinct limits of the approach: large API changes that overwhelm rule-construction capacity, and Spoon generic type constraints that block models unfamiliar with typed AST node construction.

\begin{tcolorbox}[
  enhanced,
  colback=white,
  colframe=black,
  boxrule=0.5pt,
  arc=2pt,
  left=6pt, right=6pt, top=4pt, bottom=4pt
]
\textbf{Answer to RQ1:}
\bigbag\ configurations produce compilable fixing transformations for the majority of breaking dependency updates in the benchmark, with compilation rates ranging from 50.3\% to 94.3\% across the eight model-engine combinations.
Shortfalls stem from two failure modes: \emph{Transformation Bypass} outcomes, where agents bypass the AST framework, and compilation failures, in which agents produce syntactically incorrect transformations.
JavaParser consistently yields higher CRR than Spoon across all models, with the best configuration (OpenCode/DeepSeek-v3.2/JavaParser) reaching 94.3\%, thanks to an easier transformation API with no type parameters.
\end{tcolorbox}

\subsection{RQ2: Transformation Effectiveness}
\label{RQ2}
\begin{table*}[t]
  \centering
  \caption{RQ2 (Transformation Effectiveness): outcomes when applying the generated fixing transformation to the corresponding client code and then compiling / executing the test suite ($N$ per configuration shown in the \textbf{Well-formedness} row).
    \colorbox{green!15}{\strut Green rows}: successful repairs; \textsc{patch}/\textsc{minor}/\textsc{major} break down \textsc{Build Success} by semantic versioning level.
    \colorbox{red!10}{\strut Red rows}: failure modes.
    Sp\,=\,Spoon; Jp\,=\,JavaParser. $^{*}$: highest count within that row.}
  \label{tab:rq2-combined}
  \resizebox{\textwidth}{!}{%
  {\setlength{\tabcolsep}{4pt}
  \begin{tabular}{l c c c c c c c c}
    \toprule
    \textbf{Scaffold} & \multicolumn{6}{c}{\textbf{OpenCode}} & \multicolumn{2}{c}{\textbf{Gemini CLI}} \\
    \cmidrule(lr){2-7}  \cmidrule(lr){8-9}
    \textbf{Model} & \multicolumn{2}{c}{\textbf{GPT-5.4-mini}} & \multicolumn{2}{c}{\textbf{Qwen3-30B}} & \multicolumn{2}{c}{\textbf{DeepSeek-v3.2}} & \multicolumn{2}{c}{\textbf{Gemini-3.1-Pro}} \\
    \cmidrule(lr){2-3}  \cmidrule(lr){4-5}  \cmidrule(lr){6-7}  \cmidrule(lr){8-9}
    \textbf{Engine} & \textbf{Sp} & \textbf{Jp} & \textbf{Sp} & \textbf{Jp} & \textbf{Sp} & \textbf{Jp} & \textbf{Sp} & \textbf{Jp} \\
    \midrule
    Well-formedness & 113 & 141 & 79 & 85 & 126 & 148 & 103 & 84 \\
    \midrule
    \rowcolor{green!15} \textsc{Build success \textbf{(RQ2)}}
      & 14 (12.4\%)
      & 38 (27.0\%)
      & 0 (0.0\%)
      & 2 (2.4\%)
      & 5 (4.0\%)
      & 29 (19.6\%)
      & 58 (56.3\%)
      & 66 (78.6\%) \\
    \rowcolor{green!8} \hspace{0.8em}\textsc{patch}
      & 1 & 0 & 0 & 0 & 0 & 1 & 1 & 1 \\
    \rowcolor{green!8} \hspace{0.8em}\textsc{minor}
      & 8 & 22 & 0 & 2 & 5 & 16
      & \textbf{32}$^{*}$
      & 29 \\
    \rowcolor{green!8} \hspace{0.8em}\textsc{major}
      & 5 & 16 & 0 & 0 & 0 & 12 & 25
      & \textbf{36}$^{*}$ \\
    \midrule
    \rowcolor{red!10} \textsc{Client Compilation failure}
      & 80 (70.8\%)
      & 68 (48.2\%)
      & 69 (87.3\%)
      & 75 (88.2\%)
      & 80 (63.5\%)
      & 90 (60.8\%)
      & 44 (42.7\%)
      & 17 (20.2\%) \\
    \rowcolor{red!10} \textsc{Rule execution error}
      & 17 (15.0\%)
      & 10 (7.1\%)
      & 10 (12.7\%)
      & 3 (3.5\%)
      & 39 (31.0\%)
      & 8 (5.4\%)
      & 1 (1.0\%)
      & 1 (1.2\%) \\
    \rowcolor{red!10} \textsc{Test failure}
      & 2 (1.8\%)
      & 25 (17.7\%)
      & 0 (0.0\%)
      & 5 (5.9\%)
      & 2 (1.6\%)
      & 21 (14.2\%)
      & 0 (0.0\%)
      & 0 (0.0\%) \\
    \bottomrule
  \end{tabular}%
  }%
  }%
\end{table*}

In this RQ, we measure how many compilable fixing transformations from RQ1 successfully repair the client project.
By ``successfully repair'', we mean that the project compiles and tests pass after having applied the transformation.
\autoref{tab:rq2-combined} reports the outcomes; the rows directly below \textsc{Build Success} stratify it by semantic versioning level.

\textbf{Successful repairs.}
A correct fixing transformation requires three properties: correct identification of all affected API constructs, precise localization of AST nodes at each rewrite site, and complete coverage of every API call across the client project.
GeminiCLI/Gemini-3.1-Pro/JavaParser achieves the study's highest Fix Rate at 78.6\%, with 36 fixed MAJOR breaking updates and 29 fixed MINOR breaking updates.
OpenCode/GPT-5.4-mini/JavaParser and GeminiCLI/Gemini-3.1-Pro/Spoon also repair more than a quarter of the breakages they can target with their compilable transformations.
These results confirm that the approach is not restricted to straightforward API changes: it succeeds even on multi-file MAJOR updates.

The \texttt{org.liquibase:liquibase-core} update from 3.4.2 to 4.8.0 in \href{https://github.com/chains-project/bump/blob/main/data/benchmark/feb582661e77de66eadaa7550720a8751b266ee4.json}{\texttt{sabomichal/liquibase-mssql}} (MAJOR update) illustrates all three properties.
The agent invokes \texttt{mvn compile} and identifies incompatibilities across three source files.
It then consults the dependency Javadoc to locate the correct replacement signatures: \texttt{liquibase.util.StringUtils} is renamed to \texttt{StringUtil}, and \texttt{ExecutorService.getInstance()} is replaced by a scope-based API.
GeminiCLI/Gemini-3.1-Pro/JavaParser generates a fixing transformation that applies both rewrites in a single pass, as shown in Listing~\ref{lst:liquibase-diff}.
The agent uses compiler feedback to locate the incompatibility and API documentation to resolve the replacement signature, covering all three affected files in one generated transformation.
This trajectory illustrates the self-correcting loop that underpins successful repairs.
The transformation compiles on the first attempt and \texttt{mvn test} passes.

\begin{listing}[t]
\begin{lstlisting}[language=diff]
@@ AddPrimaryKeyGeneratorMSSQL.java (same pattern in CreateIndexGeneratorMSSQL.java)
-import liquibase.util.StringUtils;
+import liquibase.util.StringUtil;
 ...
-if (StringUtils.trimToNull(statement.getTablespace()) != null)
+if (StringUtil.trimToNull(statement.getTablespace()) != null)
@@ MSSQLDatabase.java
-ExecutorService.getInstance().getExecutor(this).execute(stmt);
+Scope.getCurrentScope()
+    .getSingleton(ExecutorService.class)
+    .getExecutor("jdbc", this).execute(stmt);
\end{lstlisting}
\caption{Diff produced by applying  the transformation by GeminiCLI/Gemini-3.1-Pro/JavaParser  for fixing \texttt{liquibase-core} 3.4.2$\,{\to}\,$4.8.0 in \texttt{sabomichal/liquibase-mssql}.}
\label{lst:liquibase-diff}
\end{listing}

\textbf{Dominant failure mode.}
The main reason why the generated transformations do not succeed at fixing the breakage is \textsc{Client Compilation failure}: the transformation generates a client update that does not build.
In six of eight configurations, this category accounts for 48--88\% of outcomes, indicating that most compilable transformations target the wrong constructs or miss affected call sites entirely.
The two exceptions are the GeminiCLI/Gemini-3.1-Pro configurations, which generally produce compilable code.

OpenCode/Qwen3-30B illustrates the extreme bad case: it produces 79 (Spoon) and 85 (JavaParser) compilable transformations, yet achieves Fix Rates of 0.0\% and 2.4\%.
Its \textsc{Client Compilation failure} rates of 87.3\% (Spoon) and 88.2\% (JavaParser) are the highest in the study.
Although compilable, these transformations operate on constructs unrelated to the API element in the compilation error, leaving the project unrepaired.
OpenCode/Qwen3-30B consistently generates syntactically valid AST rules, following the instruction, but does not address the core problem at hand.

\textbf{AST transformation engine.}
The choice of AST transformation engine shapes the type of failure that occurs.
JavaParser yields significantly higher Fix Rates than Spoon for three of four models: OpenCode/GPT-5.4-mini (27.0\% vs.\ 12.4\%, $p = 0.005$), OpenCode/DeepSeek-v3.2 (19.6\% vs.\ 4.0\%, $p < 0.001$), and GeminiCLI/Gemini-3.1-Pro (78.6\% vs.\ 56.3\%, $p = 0.002$).
The Spoon disadvantage concentrates in \textsc{Rule execution error}: across the three models that exhibit this failure mode, rates are substantially higher on Spoon (12.7--31.0\%) than on JavaParser (3.5--7.1\%).
This pattern suggests that Spoon's type-resolution requirements lead transformations to crash at runtime.
\textsc{Test failure} accounts for 17.7\% of OpenCode/GPT-5.4-mini/JavaParser and 14.2\% of OpenCode/DeepSeek-v3.2/JavaParser outcomes, indicating those transformations resolve the compilation error but introduce behavioral changes in method semantics.

\textbf{Semantic versioning level.}
MINOR updates yield a higher Fix Rate than MAJOR in six of eight configurations, consistent with both intuition and their lower average error count per update (\autoref{study_subjects}).
This trend is not universal: GeminiCLI/Gemini-3.1-Pro/JavaParser is the only configuration that produces more MAJOR repairs than MINOR.
Even PATCH updates, which by semantic versioning convention should introduce no breaking changes, are not uniformly easy to repair. Overall, all our experiments converge to the fact that all PATCH cases in our dataset do not follow SemVer.

\begin{tcolorbox}[
  enhanced,
  colback=white,
  colframe=black,
  boxrule=0.5pt,
  arc=2pt,
  left=6pt, right=6pt, top=4pt, bottom=4pt
]
\textbf{Answer to RQ2:}
Coding agents can generate transformations that can verifiably repair breaking dependency updates.
A strong frontier model (Gemini-3.1-Pro) with a simple transformation API surface (JavaParser) yields the highest performance.
Fix Rates highly vary from 0.0\% to 78.6\%, and the dominant symptom of failure across configurations is \textsc{Client Compilation failure}: compilable transformations that target wrong constructs or miss affected call sites.
Practitioners deploying agentic breaking update repair should prioritize model selection and engine pairing as the primary levers for maximizing repair success.
\end{tcolorbox}

\subsection{RQ3: Transformation Generalizability}
\label{RQ3}

GeminiCLI/Gemini-3.1-Pro/JavaParser is the best-performing configuration in RQ2, achieving a Fix Rate of 78.6\% and producing 66 fixing transformations yielding a Build~Success.
Of these 66, we discard 50 for which the eligibility and verification criteria described in \autoref{meth:rq3} could not be satisfied, either because no other client reproducing the same compilation error was found, or because the number of reproduced clients was insufficient to assess generalizability.
The remaining 16 transformations originate from 14 distinct projects in Bump covering 10 libraries, spanning domains from logging and serialization to payment infrastructure and gaming APIs, providing a diverse basis for the transfer analysis.

\autoref{tab:rq3-full} reports the generalizability outcome for each of the 16 transformations, sorted by decreasing number of distinct API change types.
Each row corresponds to one breaking update; \emph{Reproduced Clients} counts the verified transfer targets for that update (totalling 129, as identified in \autoref{meth:rq3}); \emph{Fixed Clients} counts those where the transformation repairs the compilation error; and \emph{Fix Rate} is their ratio.
The 129 reproduced clients are unevenly distributed across the 16 cases, ranging from 2 to 31 clients per update.

\begin{table*}[t]
  \centering
  \caption{RQ3 (Transformation Generalizability): cross-project transfer
    results for the 16 Gemini-3.1-Pro/JavaParser fixing transformations,
    sorted by decreasing number of distinct API change types.
    \emph{Reproduced Clients}\,=\,external repos where the old dependency
    version is declared and the broken API element appears in source
    (verified transfer targets).
    \emph{Fixed Clients}\,=\,repos achieving \texttt{mvn test-compile}
    after transformation application; $^{*}$\,=\,at least one Fixed repo
    also passes \texttt{mvn test}.
    \emph{Fix Rate} (Cross-Project Fix Rate)\,=\,Fixed\,/\,Reproduced Clients.
    Full commit hashes are available in the replication package.}
   \label{tab:rq3-full}
  \resizebox{\textwidth}{!}{%
  \begin{tabular}{c l l c l r r r r}
    \toprule
    \textbf{BU} & \textbf{Library} & \shortstack[r]{\textbf{Breaking}\\\textbf{Update}} & \textbf{SV}
      & \shortstack[l]{\textbf{BUMP}\\\textbf{Project}}
      & \shortstack[r]{\textbf{API}\\\textbf{Changes}}
      & \shortstack[r]{\textbf{Reproduced}\\\textbf{Breakages}}
      & \shortstack[r]{\textbf{Fixed}\\\textbf{Clients}}
      & \shortstack[r]{\textbf{Fix}\\\textbf{Rate}} \\
    \midrule
    \href{https://github.com/chains-project/bump/blob/main/data/benchmark/ea03f6488449fcfe8cd0a678b4c64891e1427a32.json}{BU01} & \texttt{jedis} & 3.9.0\,$\to$\,4.3.2 & major & \texttt{JRedisGraph} & 8 & 2 & 0 & 0\% \\
    \href{https://github.com/chains-project/bump/blob/main/data/benchmark/402082609522c66a3b790aedafd0570148a7d53f.json}{BU02} & \texttt{spongeapi} & 7.4.0\,$\to$\,8.0.0 & major & \texttt{ChangeSkin} & 5 & 2 & 2\rlap{$^{*}$} & 100\% \\
    \href{https://github.com/chains-project/bump/blob/main/data/benchmark/1a2fb9f65e12d6c43a8472b9b035299b29a75ce8.json}{BU03} & \texttt{jakarta.validation-api} & 2.0.2\,$\to$\,3.0.1 & major & \texttt{wicket-crudifier} & 3 & 5 & 5\rlap{$^{*}$} & 100\% \\
    \href{https://github.com/chains-project/bump/blob/main/data/benchmark/9218cc9c8e0018d01e2d7cfe0e77aae7b65b378f.json}{BU04} & \texttt{jakarta.validation-api} & 2.0.2\,$\to$\,3.0.2 & major & \texttt{wicket-crudifier} & 3 & 15 & 11\rlap{$^{*}$} & 73\% \\
    \href{https://github.com/chains-project/bump/blob/main/data/benchmark/3f30dfff617fd652412260ecf648a25769a27101.json}{BU05} & \texttt{jakarta.interceptor-api} & 1.2.5\,$\to$\,2.0.0 & major & \texttt{cdi-test} & 2 & 5 & 1\rlap{$^{*}$} & 20\% \\
    \href{https://github.com/chains-project/bump/blob/main/data/benchmark/4bba3fb6147e72946f64724fe55eee5d15ff6206.json}{BU06} & \texttt{jakarta.annotation-api} & 1.3.5\,$\to$\,2.0.0 & major & \texttt{cdi-test} & 2 & 15 & 1 & 7\% \\
    \href{https://github.com/chains-project/bump/blob/main/data/benchmark/b1a941400d68445d76056ab8833cd6d2e3455954.json}{BU07} & \texttt{snakeyaml} & 1.33\,$\to$\,2.0 & major & \texttt{fluxtion} & 2 & 2 & 0 & 0\% \\
    \href{https://github.com/chains-project/bump/blob/main/data/benchmark/d401e189fb6435110e3dc4ca1a94838f167e7ddf.json}{BU08} & \texttt{logback-classic} & 1.2.11\,$\to$\,1.4.4 & minor & \texttt{pdb} & 2 & 6 & 0 & 0\% \\
    \href{https://github.com/chains-project/bump/blob/main/data/benchmark/90ffd2cd31edecf778d14d0015da9ceab7e53081.json}{BU09} & \texttt{logback-classic} & 1.2.11\,$\to$\,1.4.0 & minor & \texttt{pay-adminusers} & 2 & 8 & 0 & 0\% \\
    \href{https://github.com/chains-project/bump/blob/main/data/benchmark/07fad972bb884e9fa6143b4f870d08305811607d.json}{BU10} & \texttt{logback-classic} & 1.2.11\,$\to$\,1.4.1 & minor & \texttt{pay-adminusers} & 2 & 3 & 0 & 0\% \\
    \href{https://github.com/chains-project/bump/blob/main/data/benchmark/0a11c04038eae517540051dbf51f7f26b7221f20.json}{BU11} & \texttt{snakeyaml} & 1.24\,$\to$\,2.0 & major & \texttt{simplelocalize-cli} & 2 & 3 & 1\rlap{$^{*}$} & 33\% \\
    \href{https://github.com/chains-project/bump/blob/main/data/benchmark/54857351e0b0a655970d7e2ccdb67f175cc5d688.json}{BU12} & \texttt{zip4j} & 1.3.2\,$\to$\,2.10.0 & major & \texttt{allure-maven} & 1 & 31 & 11\rlap{$^{*}$} & 35\% \\
    \href{https://github.com/chains-project/bump/blob/main/data/benchmark/88c1f903cede03ff371059cdaf009dab12007043.json}{BU13} & \texttt{zip4j} & 1.3.2\,$\to$\,2.11.0 & major & \texttt{allure-maven} & 1 & 24 & 8\rlap{$^{*}$} & 33\% \\
    \href{https://github.com/chains-project/bump/blob/main/data/benchmark/16ae40b1e17e14ee3ae20ac211647e47399a01a9.json}{BU14} & \texttt{zip4j} & 1.3.2\,$\to$\,2.11.1 & major & \texttt{allure-maven} & 1 & 4 & 3\rlap{$^{*}$} & 75\% \\
    \href{https://github.com/chains-project/bump/blob/main/data/benchmark/cd5bb39f43e4570b875027073da3d4e43349ead1.json}{BU15} & \texttt{plexus-utils} & 3.5.1\,$\to$\,4.0.0 & major & \texttt{pgpverify-maven-plugin} & 1 & 2 & 0 & 0\% \\
    \href{https://github.com/chains-project/bump/blob/main/data/benchmark/923a6b2027e3ca1762deb6a60fc0a768c284122b.json}{BU16} & \texttt{hamcrest-library} & 1.3\,$\to$\,2.2 & major & \texttt{jcabi-http} & 1 & 2 & 0 & 0\% \\
    \midrule
    \textbf{Total} & & & & & & \textbf{129} & \textbf{43} & \textbf{33.3\%} \\
    \bottomrule
  \end{tabular}%
  }
  \smallskip\\
  \raggedright\footnotesize
  $^{*}$~At least one Fixed client also passes \texttt{mvn test}; unmarked Fixed values are fixed by \texttt{mvn test-compile} only.
\end{table*}

\textbf{Overall transferability.}
Fixing transformations generated for one client regularly transfer to other clients affected by the same breaking update.
Success is concentrated in breaking updates where all clients use the affected API element in the same way.
Across 129 reproduced identical breaking updates, 43 transformations achieve a cross-project Build~Success, yielding a Cross-Project Fix Rate of 33.3\%.
Transfer is concentrated in 9 of the 16 transformations; the remaining transformations do not fix any other repositories.
For 8 of those 9, at least one fixed repository also passes the full \texttt{mvn test} suite, showing that successful transfers fully preserve test-level correctness alongside compilability.
\href{https://github.com/chains-project/bump/blob/main/data/benchmark/1a2fb9f65e12d6c43a8472b9b035299b29a75ce8.json}{BU03} and \href{https://github.com/chains-project/bump/blob/main/data/benchmark/9218cc9c8e0018d01e2d7cfe0e77aae7b65b378f.json}{BU04} (\texttt{jakarta.validation-api}) illustrate this success case: together they achieve 80\% across 20 reproduced clients by renaming usages of \texttt{javax\allowbreak.validation.*} to \texttt{jakarta\allowbreak.validation.*}, a uniform substitution that applies to any client regardless of how the package is invoked.
\href{https://github.com/chains-project/bump/blob/main/data/benchmark/402082609522c66a3b790aedafd0570148a7d53f.json}{BU02} (\texttt{spongeapi}) further illustrates this: despite covering five API change types, it fixes both reproduced clients (100\%) because both use the affected elements in the same way as the seed.

\textbf{Transfer failure analysis.}
Let us now discuss the failure cases.
The five transformations covering a single API change type (BU12--BU16) achieve only 35\% (22 of 63 reproduced clients), showing that transformation complexity alone does not explain failure.

Inspection of the 86 non-fixed build logs reveals two failure patterns.
In 72 of the 86 failures, the rule runs without error but finds no call sites to transform: the matching pattern of the transformation is incorrect.
In 14 cases, the matching pattern is too loose.
\href{https://github.com/chains-project/bump/blob/main/data/benchmark/4bba3fb6147e72946f64724fe55eee5d15ff6206.json}{BU06} (\texttt{jakarta.annotation-api}) is exemplar: the transformation correctly renames JSR-250 annotations (\texttt{javax.annotation.*} $\to$ \texttt{jakarta.annotation.*}), as the seed uses, but applies the same rename to \texttt{@Nullable} (a JSR-305 annotation absent from \texttt{jakarta.annotation-api:2.0.0}), introducing a new compilation error in 14 of 15 transfer targets.

Together, these two patterns illustrate the fundamental constraint: because transformation generation is conditioned by a single project's compilation errors, the rule accurately repairs clients that reach the broken API through the same code path as the seed, but leaves unobserved patterns in other clients unrepaired.
Possible directions to overcome this problem include seeding the agent with multiple client projects to expose a broader range of usage patterns, or providing the complete list of changed API elements rather than deriving the rewrite scope from a single project's compilation errors.

\begin{tcolorbox}[
  enhanced,
  colback=white,
  colframe=black,
  boxrule=0.5pt,
  arc=2pt,
  left=6pt, right=6pt, top=4pt, bottom=4pt
]
\textbf{Answer to RQ3:}
Fixing transformations to repair breaking updates transfer across projects, achieving an overall Cross-Project Fix Rate of 33.3\% (43 of 129 verified targets), with some generalization demonstrated in 9 of 16 breaking updates.
The failures are mostly due to the single-project seed: in this failure mode, the agent encodes only the usage patterns visible in one project's compilation errors, failing to generalize to all projects from one single example.
\end{tcolorbox}

\section{Threats to Validity}
\label{sec:threats}

\textbf{Construct validity.}
Our metrics measure observable build outcomes per \texttt{mvn test} success, which does not guarantee the absence of behavioral regressions.
The conservative transfer criterion (accepting only clients that reference the same broken API element as the seed project) may exclude clients that would benefit from the transformation but manifest the breaking change through a different compilation error.

\textbf{Internal validity.}
Each (model, engine, breaking update) combination is executed once; LLM nondeterminism~\cite{bjarnason2026randomness} means individual runs may deviate from average behavior.
The consistency of results across all four models within the same engine, and across both engines, makes it unlikely that nondeterminism drives the observed patterns.

\textbf{External validity.}
The study is scoped to Java projects built with Maven, using two Java-specific AST transformation engines; generalizability to other languages or build systems cannot be assumed without further evaluation.
The 157 breaking updates span 69 client projects and 70 libraries across diverse domains, but are restricted to open-source, reproducible compilation failures; results may not generalize to closed-source or industrial codebases.
The transfer analysis seeds each transformation from a single client project, which may underrepresent the diversity of API usage patterns in the broader ecosystem.

\textbf{Conclusion validity.}
Pairwise statistical comparisons use Fisher's exact test; with three pairwise dimensions (model, engine, SemVer level) across eight configurations, multiple-comparison inflation is possible.
We restrict causal claims to effects that are large and consistent across both engines, reducing the risk of false positives.

\section{Related Work}
\label{sec:rw}

\subsection{API Evolution and Breaking Dependency Updates}
\label{rw:api-evolution}

Prior work on API evolution has characterised and detected breaking changes at scale.
Dig and Johnson~\cite{dig2006apis} establish a foundational taxonomy of API-breaking changes, classifying them into structural refactoring operations (method moves, renames, and signature changes) and behavioral modifications. Across five Java components, they find that over 80\% of breaking changes are refactorings, with method moves and renames as the dominant types.
Ochoa~\etal~\cite{ochoa2022breaking} confirm that semantic versioning alone is insufficient to protect downstream Maven clients, with a small fraction of high-usage libraries responsible for a disproportionate share of ecosystem-wide breakage.
Brito~\etal~\cite{8330214} establish a taxonomy of breaking-change categories in open-source Java; removal of deprecated elements and signature modifications are the dominant types, directly informing the AST transformations a repair agent must support.

Closer to our setting, several works address migration assistance and rule mining.
Dagenais and Robillard~\cite{dagenais2011recommending} propose SemDiff, which recommends adaptive changes by mining framework-client co-evolution histories; it requires prior migration data that is unavailable for first-occurrence breaking updates.
Reyes~\etal~\cite{Reyes2024BreakingGood} analyze 243 real-world breaking dependency updates from BUMP using build-log and dependency-tree analysis, identifying root causes for 70\% of them. Their taxonomy of compilation error categories directly characterizes the failure classes \bigbag\ targets, providing explanations to guide developers rather than automated repair.
Ramos~\etal~\cite{Ramos2023MELT} mine 461 correct API-migration rules from 1179 pull requests across four Python libraries using syntax-driven find-and-replace patterns in Comby. Their approach requires existing migration examples in library pull requests and produces lightweight syntactic patterns rather than fixing transformations applicable to previously unseen breaking changes.

More recent work has extended the scope from detection to automated repair.
Fruntke and Krinke~\cite{FruntkeLukas2025} evaluate zero-shot prompting and an iterative LLM agent on the BUMP dataset, showing that LLMs can address this failure class without historical migration data.
Reyes~\etal~\cite{reyes2025byam} demonstrate that contextual LLM prompts repair up to 27\% of breaking dependency updates, but generate code patches rather than transferable fixing transformations.
On the industrial side, FOSSA~\cite{fossa} automatically reviews dependency updates and analyses their code impact, while Patchwork~\cite{patchwork} goes further by comparing dependency versions, generating patches, and validating fixes within a continuous-integration pipeline.
Our work produces structured, transferable fixing transformations; a single fixing transformation derived from one affected client transfers to any other client broken by the same dependency update, a property that patch-based approaches do not offer.

\subsection{Automated Program Repair}
\label{rw:apr}

Automated program repair (APR) has evolved from search-based techniques~\cite{le2011genprog} to LLM-driven strategies~\cite{xia2023automated,sobania2023analysis}, but existing evaluations focus on general bug corpora and single-turn generation, without addressing failures caused by dependency updates.
Le~Goues~\etal~\cite{le2011genprog} demonstrate that genetic-programming-guided edits can automatically repair real-world bugs, but their approach requires a passing test suite, which is unavailable in the compilation-failure setting due to dependency updates.
Rolim~\etal~\cite{Sousa2016LearningSP} show that program transformations can be synthesized from code-edit examples (REFAZER), but require developer-provided before-and-after demonstrations unavailable for first-occurrence breaking dependency updates.

LLM-based approaches have since substantially raised the bar on standard APR benchmarks.
Xia~\etal~\cite{xia2023automated} show that pre-trained LLMs now outperform traditional generate-and-validate approaches on standard APR benchmarks.
Sobania~\etal~\cite{sobania2023analysis} evaluate ChatGPT on Defects4J in a single-turn setting and note that interactive multi-turn refinement could substantially improve outcomes.
Tarlow~\etal~\cite{tarlow2020learning} train Graph2Diff on 500,000 professional build errors to predict AST diffs that fix compilation failures, but produce instance-level patches without cross-project transferability.
Fu~\etal~\cite{Fu2025AutoRepair} show that LLMs with fix templates can repair compilation errors without a test suite, achieving a 63\% CI pass rate across 1,000 industrial C/C++ build failures.
Our work addresses the same repair setting but frames it as the generation of executable AST processors; the produced fixing transformations are transferable to any client affected by the same breaking update, not just the instance under repair.

\subsection{Agentic Code Generation}
\label{rw:llm-agents}

LLM-based agentic frameworks have shown strong results on real-world software tasks, yet none has examined how the choice of target transformation library affects repair effectiveness.
LLMs exhibit strong code generation capability~\cite{chen2021evaluating,brown2020language} and, when combined with tool-augmented agentic loops, can resolve real-world repository tasks~\cite{jimenez2023swe}.
Chen~\etal~\cite{chen2021evaluating} demonstrate that large-scale code pre-training yields strong zero-shot synthesis, underpinning the assumption that frontier models carry implicit API knowledge.
Brown~\etal~\cite{brown2020language} show that in-context examples alone can steer frontier model behavior without parameter updates.
Jimenez~\etal~\cite{jimenez2023swe} establish, using SWE-bench, that resolving multi-file real-world tasks demands agentic, multi-turn LLM behavior; the agent-computer interface paradigm that enables this is now well established~\cite{yang2024swe}.

Single-pass LLM generation, however, struggles with version-conditioned tasks and complex dependency migrations.
Misra~\etal~\cite{Misra2025GitChameleon} show that frontier LLMs achieve only 48--51\% success on 328 version-conditioned Python problems, revealing the limits of single-pass generation and motivating \bigbag's iterative, feedback-driven repair approach.
May~\etal~\cite{May2025FreshBrew} show on 228 Java project-level migrations that rule-based tools cannot handle unforeseen dependency incompatibilities (7\% success), a gap \bigbag\ addresses with agent-synthesised fixing transformations.

Our work applies this agentic paradigm to compilation error repair and specifically shows that the choice of target AST transformation engine is a first-order determinant of repair success.

\subsection{LLM-Generated Transformation Synthesis}
\label{rw:transf-synthesis}

A direct line of work uses LLMs to generate reusable transformations rather than one-off patches, which is precisely the paradigm \bigbag\ instantiates for breaking dependency updates.

SPELL~\cite{Ramos2026SPELL} extends MELT~\cite{Ramos2023MELT} by eliminating the migration-corpus requirement: it prompts an LLM to generate synthetic migration examples compiled into PolyglotPiranha transformation scripts, validated on sibling implementations (63.3\% migration rate) and applied to 18 Python projects across ten migration tasks.
Allain~\etal~\cite{allain2026code} systematically evaluate LLMs on generating transformation rules expressed in three DSLs (Comby, Ast-Grep, GritQL) across six datasets; frontier models achieve high syntactic validity but accuracy degrades sharply for multi-statement migrations where localized DSL patterns fail to capture the full repair scope.
Dilhara~\etal~\cite{Dilhara2024PyCraft} combine LLMs with transformation-by-example: given developer-provided examples of a code change pattern, PyCraft generates semantic variants and synthesizes reusable ComBy rules; however, it requires pre-existing change examples and targets known, repetitive patterns.
Cummins~\etal~\cite{Cummins2024DontTransform} demonstrate that synthesizing executable Python \texttt{ast.AST} transformation functions from input/output examples outperforms direct code rewriting when combined with iterative sandboxed execution feedback.

\bigbag\ shares with all four the core decision to generate reusable transformation programs rather than patches, and extends this paradigm along three axes none of them cover jointly: it targets breaking updates; it synthesizes type-aware transformation programs via Spoon and JavaParser rather than syntactic DSL scripts or Python AST functions; and it proposes a novel and sound methodology to study generalizability.

\subsection{Dependency Update Automation}
\label{rw:dep-automation}

Automated dependency-management tools address update discovery and notification~\cite{mirhosseini2017can,kula2018developers} but do not repair source-level compilation breakage introduced by a breaking update.
Mirhosseini and Parnin~\cite{mirhosseini2017can} show that automated pull requests achieve acceptance rates competitive with manually submitted ones, demonstrating the viability of bot-driven update nudges.
Kula~\etal~\cite{kula2018developers} find that most developers leave dependencies outdated, naming the absence of automated source-repair support as a primary barrier to timely migration.

Our work is complementary to these efforts: our agent automates the downstream source-repair step that tools such as Dependabot and Renovate leave to the developer.

\section{Conclusion}
\label{sec:conclusion}
In this paper, we have presented \bigbag, a framework that generates structured, executable, cross-project fixing transformations for breaking dependency updates using a coding agent.
We have evaluated \bigbag on \data{} breaking dependency updates across eight configurations combining four LLMs and two AST transformation engines.
Our experimental results show that coding agents produce compilable fixing transformations across most configurations: the best configuration (DeepSeek-v3.2/JavaParser) reaches a Compilable Rule Rate of 94.3\%, and the best repair configuration (GeminiCLI/Gemini-3.1-Pro/JavaParser) achieves a Fix Rate of 78.6\%.
Both model and transformation engine choice substantially affect effectiveness.
A cross-project transfer analysis over 129 verified transfer targets yields a Cross-Project Fix Rate of 33.3\% overall, rising to 80\% or above for breaking updates where all clients invoke the affected API element uniformly.

As future work, we want to improve generalizability by seeding the agent with multiple client projects.
This is important to ensure that the generated transformations capture the full usage surface of a broken API, not just the patterns present in the seed.

\bibliographystyle{IEEEtran}
\balance
\bibliography{references}

\end{document}